%% file: main.tex
\PassOptionsToPackage{dvipsnames}{xcolor}

\documentclass[twocolumn]{aastex701}  

\input{abbrev}


\input{code}

\input{concepts}

\input{derivatives}

\input{formatting}

\input{nuclides}

\input{symbols}
\input{units}
\input{vectors}

\usepackage{soul}

\shorttitle{O-C Mergers and $\mathsf{^{44}}$Ti}
\begin{document}

\title{Pre-supernova O-C shell mergers could produce more $\mathsf{^{44}}$Ti than the explosion}

\author[0000-0002-1283-6636,gname=Joshua,sname=Issa]{Joshua Issa}
\affiliation{Astronomy Research Centre, Department of Physics \& Astronomy, University of Victoria, Victoria, BC, V8W 2Y2, Canada}
\affiliation{NuGrid Collaboration, \url{http://nugridstars.org}}
\email[show]{joshuaissa@uvic.ca}  

\correspondingauthor{Joshua Issa}

\author[0000-0001-8087-9278,gname=Falk,sname=Herwig]{Falk Herwig}
\affiliation{Astronomy Research Centre, Department of Physics \& Astronomy, University of Victoria, Victoria, BC, V8W 2Y2, Canada}
\affiliation{NuGrid Collaboration, \url{http://nugridstars.org}}
\email{fherwig@uvic.ca}

\received{January 26, 2026}
\revised{April 24, 2026}
\accepted{May 11, 2026}

\submitjournal{ApJ}

\begin{abstract}
    The formation of $^{44}\mathrm{Ti}$ in massive stars is thought to occur during explosive nucleosynthesis, however recent studies have shown it can be produced during O-C shell mergers prior to core collapse. We investigate how mixing according to 3D macro physics derived from  hydrodynamic simulations impacts pre-supernova O-C shell merger nucleosynthesis and if it can dominate explosive supernova production of $^{44}\mathrm{Ti}$ and other radioactive isotopes. We compare a range of observations and models of explosive $^{44}\mathrm{Ti}$ yields to pre-explosive multi-zone mixing-burning nucleosynthesis simulations of an O-C shell merger in a $15~\mathrm{M_\odot}$ $Z=0.02$ stellar model with mixing conditions corresponding to different 3D hydro mixing scenarios. Radioactive species produced in the $\mathrm{O}$ shell have a multi-dex spread in pre-explosive yield predictions across different 3D mixing scenarios of $1.54~\mathrm{dex}$ and $2.14~\mathrm{dex}$ on average depending on mass cut. $^{44}\mathrm{Ti}$ has the largest spread of $4.78~\mathrm{dex}$ and $4.81~\mathrm{dex}$ depending on mass cut. Further, we show that the pre-explosive production of $^{44}\mathrm{Ti}$ can be larger than the explosive production of models and can match observations. Our results also show that 3D mixing physics enhances $^{44}\mathrm{Ti}$ in 1D models without modifying $^{56}\mathrm{Ni}$ yields. We conclude that quantitative predictions of $^{44}\mathrm{Ti}$ and other radioactive species more broadly require an understanding of the 3D hydrodynamic mixing conditions present during the O-C shell merger.
\end{abstract}


\section{Introduction}

The radioactive isotope \titanium[44] is a key tracer of explosive nucleosynthesis in core collapse supernovae (CCSN). 
It is primarily synthesized during $\alpha$-rich freeze-out from nuclear statistical equilibrium in the innermost ejecta associated with explosive \silicon[] burning by the reactions \calcium[40]$(\alpha,\gamma)$ and \scandium[43]$(\pt,\gamma)$ and co-produced with \nickel[56] \citep{woosleyEvolutionExplosionMassive1995, theAre44TiproducingSupernovae2006, magkotsiosTRENDS44Ti56Ni2010, sieverdingProduction44TiIrongroup2023}.
It is observed by the emission lines $67.87$, $78.32$, and $\unit{1157}{\kilo\eV}$ from the decay chain $\titanium[44] \rightarrow \scandium[44] \rightarrow \calcium[44]$ in supernova remnants \citep{iyudinCOMPTELObservations44Ti1994, grebenevHardXrayEmissionLines2012, boggs44TiGammarayEmission2015, grefenstetteAsymmetriesCorecollapseSupernovae2014, grefenstetteDistributionRadioactive44Ti2017}. 

Constraining the explosive origin of \titanium[44] in supernova remnants is challenging because the ejected yield is sensitive to the mass cut and its co-produced \nickel[56].
Models must reproduce both simultaneously to be physically consistent, but 1D models struggle to do this without overproducing \nickel[56] \citep{chieffiProduction26Al60Fe2002,chieffiSynthesis44Ti56Ni2017}. 
3D explosion models are required to match observations since spherically symmetric explosions underproduce the \titanium[44] yields inferred from observations of Cassiopeia A and SN 1987A \citep{wongwathanaratProductionDistribution44Ti2017,sieverdingProduction44TiIrongroup2023,wangInsightsProduction44Ti2024}.

Recently it has been shown that \titanium[44] can be significantly produced prior to the explosion during an \oxygen[]-\carbon[] shell merger in massive stars \citep{turPRODUCTION26Al44Ti2010, chieffiSynthesis44Ti56Ni2017}, and that the pre-explosive yield can partially survive the explosion \citep{robertiOccurrenceImpactCarbonoxygen2025a}. 
Mergers occur a few hours prior to the CCSN of a red supergiant, therefore \titanium[44] ($t_{1/2}=\unit{59.1}{\yr}$) synthesized in this event can contribute to the observed signal seen in supernova remnants.
This may be the case for Cassiopeia A, as observations of its \silicon[], \neon[], \chlorine[], and \phosphorus[] have recently been interpreted as signatures of an \oxygen[]-\carbon[] shell merger prior to collapse \citep{satoInhomogeneousStellarMixing2025, audardChlorinePotassiumEnrichment2025}.

Cassiopeia A \citep{iyudinCOMPTELObservations44Ti1994, grefenstetteAsymmetriesCorecollapseSupernovae2014, grefenstetteDistributionRadioactive44Ti2017} and SN 1987A \citep{grebenevHardXrayEmissionLines2012, boggs44TiGammarayEmission2015} both have asymmetric and clumpy distributions of \titanium[44].
This may be a signature of the pre-explosive origin of \titanium[44] in an \oxygen[]-\carbon[] shell merger, as 3D hydrodynamic simulations show mergers have asymmetric and large-scale non-radial flows whose features could be preserved in the remnant \citep{andrassy3DHydrodynamicSimulations2020,yadavLargescaleMixingViolent2020,rizzutiShellMergersLate2024a,satoInhomogeneousStellarMixing2025}.
Nucleosynthesis during mergers in these 3D macro physical flows diverges significantly from 1D stellar evolution predictions in the \oxygen[] shell \citep{yadavLargescaleMixingViolent2020, rizzutiShellMergersLate2024a,issaImpact3DMacrophysics2026, issa3DMacroPhysics2026}.   

In this study, we explore how pre-explosive nucleosynthesis of \titanium[44] and other radioactive isotopes depend on the mixing conditions during an \oxygen[]-\carbon[] shell merger as motivated by 3D hydrodynamic simulations.
This paper is organized as follows: \Sect{methods} provides an overview of the post-processed O-C shell conditions, \Sect{results} demonstrates how \titanium[44] is produced non-explosively and compares the production in the \oxygen[]-\carbon[] shell merger to explosive yields, and \Sect{conclusions} discusses the observational implications of the results.

\section{Methods}\label{sec:methods}

The NuGrid data set includes massive star models with initial masses $12{-}\unit{25}{\Msun}$ and initial metallicities $Z = 10^{-4}$ to $0.02$ \citep{ritterNuGridStellarData2018}. 
These models were calculated in 1D using \MESA{} \citep{paxtonMODULESEXPERIMENTSStelLAR2010} without rotation and no convective overshooting after the end of core helium burning \citep{pignatariNuGridStellarData2016,ritterNuGridStellarData2018}. 
Five models have an \oxygen[]-\carbon[] shell merger shortly before core collapse: $\unit{15}{\Msun}$ $Z=0.02$, $\unit{12}{\Msun}$ $Z=0.01$, $\unit{15}{\Msun}$ $Z=0.01$, $\unit{20}{\Msun}$ $Z=0.01$, and $\unit{12}{\Msun}$ $Z=0.001$.
Not all shell mergers feature high levels of pre-explosive \titanium[44] production due to different locations of merger, temperatures, sizes, and time before core collapse, but two of these models have this feature (see \Tab{ti_ritter} in \Appendix{appendix_explosive_yields}).

\begin{figure*}[!ht]
\centering
\includegraphics[width=\hsize]{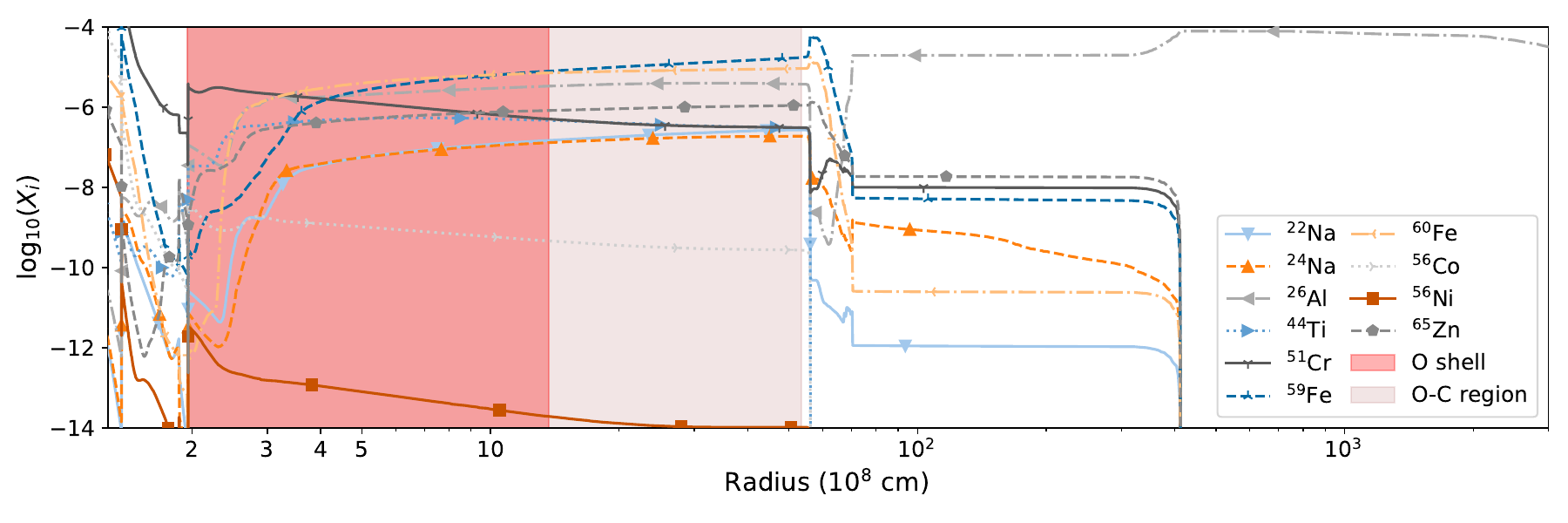}
    \caption{Mass fractions of radioactive isotopes in the $\unit{15}{\Msun}$ $Z=0.02$ model from~\cite{ritterNuGridStellarData2018} at the final time step before explosion. 
    Both the \oxygen[] shell and \oxygen[]-\carbon[] region are shaded to indicate their extent.
    The mass cut for the rapid and delayed explosion prescriptons are located at \unit{\natlog{1.3}{8}}{\cm} (the left edge) and \unit{\natlog{3.7}{8}}{\cm} (within the \oxygen[] shell) respectively.}
\label{fig:massfrac}
\end{figure*}

We have post-processed the initial onset of a merger with the \oxygen[] shell of the $\unit{15}{\Msun}$ $Z=0.02$ NuGrid multi-zone code \mppnp{} \citep{pignatariNuGridStellarData2016} to study how nucleosynthesis of the light odd-Z isotopes and \pnucn{} are changed by 3D-inspired mixing scenarios \citep{issaImpact3DMacrophysics2026,issa3DMacroPhysics2026}. 
This stellar model produces a number of radioactive species in the \oxygen[]-\carbon[] merger prior to the explosion as shown in \Fig{massfrac}.

The merger is simulated for \unit{110}{\second}, using mass fractions shortly prior to the merger and a static stellar structure from the onset with a network of 1470 isotopes.
Full star yields are estimated by modifing the pre-explosive yields of~\cite{ritterNuGridStellarData2018} in the shell merger region by the ratio with our results at \unit{110}{\second} as described in~\cite{issa3DMacroPhysics2026}.
This neglects feedback effects that require full stellar evolution as reactions will change the temperature and density evolution.
Therefore, we caution that these yields are exploratory rather than predictive.

We present 24 simulations with different mixing profiles and entrainment rates. 
One mixing profile is used from the \MESA{} calculations according to mixing length theory (MLT), and the rest are modified according to results seen in 3D hydrodynamic simulations \citep[Figures~4 and 5,][]{issaImpact3DMacrophysics2026}.
Following the prescription of \cite{jonesIdealizedHydrodynamicSimulations2017} we implement a convective downturn and boosting with the following equation:
\[
   D_{\mathrm{3D{-}insp.}} = \frac{1}{3} v_{\mathrm{MLT}} \times \min(\ell,|r-r_0|)
\]
where $v_{\mathrm{MLT}}$ are the velocities predicted by MLT, $\ell$ is the mixing length, and $|r-r_0|$ is the radial distance to the convective boundary that enables a gradual downturn.
We boost the mixing efficiency by factors of $1$, $3$, $10$, and $50$ to this downturn profile \citep[a range covering what is seen in the radial velocities of 3D simulations:][]{jonesIdealizedHydrodynamicSimulations2017,andrassy3DHydrodynamicSimulations2020,rizzutiShellMergersLate2024a}. 
The MLT and downturn scenarios are calculated with different \carbon[]-shell entrainment rates of 0, \natlog{4}{-5}, \natlog{4}{-4}, and \unit{\natlog{4}{-3}}{\Msun\sec^{-1}}, as the simulations of~\cite{andrassy3DHydrodynamicSimulations2020} show that entrainment depends on the luminosity of $\carbon[12]+\carbon[12]$ burning.
Four profiles with convective dips emulating quenching effects due to possible strong energy feedback and partial merging \citep[as suggested by][]{andrassy3DHydrodynamicSimulations2020} are also included and calculated with a single entrainment rate of \unit{\natlog{4}{-3}}{\Msun\sec^{-1}}.

We provide results in terms of an overproduction factor defined as
\[
\OP_{\mathrm{R}15}=X_{\mathrm{model}}/X_{\mathrm{Pre-Explosive~\unit{15}{M_\odot}~Z=0.02}}
\]
where $X_{\mathrm{model}}$ is the ejected yield for a species from a given model and $X_{\mathrm{Pre-Explosive~\unit{15}{M_\odot}~Z=0.02}}$ is the undecayed pre-explosive yield from the $\unit{15}{\Msun}$ $Z=0.02$ model from \cite{ritterNuGridStellarData2018}.
We provide yields for both rapid and delayed mass cuts used in the NuGrid data set, following \cite{fryerCOMPACTREMNANTMASS2012}, which provides a semi-analytical prescription of the compact mass remnant after explosion and fallback.
Further details about these simulations and yield estimation can be found in~\cite{issaImpact3DMacrophysics2026, issa3DMacroPhysics2026}.

\subsection{Comparison to Observations}\label{sec:obs_citations}

We compare our results to observations of \titanium[44] in supernova remnants.
We include observations of the core-collapse remnants Cassiopeia A \citep{iyudinCOMPTELObservations44Ti1994, vinkDetection6797842001, siegertRevisitingINTEGRALSPI2015, wangHardXRayEmissions2016, grefenstetteDistributionRadioactive44Ti2017, weinberger44TiEjectaYoung2020} with an ejecta mass of \titanium[44] between \unit{\natlog{(0.8{-}2.6)}{-4}}{\Msun} and core-collapse remnant SN 1978A \citep{jerkstrand44TipoweredSpectrumSN2011, grebenevHardXrayEmissionLines2012,boggs44TiGammarayEmission2015, weinberger44TiEjectaYoung2020} with an ejecta mass of \titanium[44] between \unit{\natlog{(1.5{-}6.9)}{-4}}{\Msun}.
These studies provide integrated yields except the spatially resolved yields of \cite{grefenstetteDistributionRadioactive44Ti2017}.

\subsection{Comparison to Models}\label{sec:model_citations}

We also compare our results to a number of multi-D models from the literature with \titanium[44] yields. 
\cite{woosleyEvolutionExplosionMassive1995} have 1D models that yield $\natlog{1.35}{-33}{-}\unit{\natlog{2.44}{-4}}{\Msun}$ of \titanium[44] for massive stars with initial masses of \unit{11{-}40}{\Msun} and metallicites \unit{0{-}1}{Z_\odot}.
\cite{magkotsiosTRENDS44Ti56Ni2010} have a 1D supernova and hypernova of a \unit{16}{\Msun}, $Z_\odot$ star with yields \unit{\natlog{1.04}{-4}}{\Msun} and \unit{\natlog{2.66}{-5}}{\Msun}, and a rotating 2D supernova of a {15}{\Msun}, $Z_\odot$ star that yields \unit{\natlog{6.98}{-5}}{\Msun}.
\cite{sukhboldCorecollapseSupernovae92016} has a set of massive stars of \unit{9{-}120}{\Msun}, $Z_\odot$ with 1D neutrino-driven supernovae that yield up to \unit{\natlog{7.45}{-5}}{\Msun}, some of which have been idenitfied to have \oxygen[]-\carbon[] shell mergers \citep{satoInhomogeneousStellarMixing2025}.
\cite{chieffiSynthesis44Ti56Ni2017} have both 1D non-rotating and rotating models of \unit{13{-}120}{\Msun} and metallicities $[\iron[]/\hydrogen[]]=-1, 0$ that yield \unit{\natlog{(0.6{-}6.0)}{-5}}{\Msun}. 
13 of these models were identified by \cite{robertiOccurrenceImpactCarbonoxygen2025a} to have \oxygen[]-\carbon[] shell mergers.
\cite{eichlerNucleosynthesis2DCorecollapse2017} have a 2D supernova of a \unit{17}{\Msun}, $Z_\odot$ that yields \unit{\natlog{(1.14{-}1.35)}{-5}}{\Msun}.
\cite{wongwathanaratProductionDistribution44Ti2017} have a 3D neutrino-driven supernova of a \unit{15}{\Msun}, $Z_\odot$ that yields \natlog{8.66}{-6}${-}$\unit{\natlog{1.49}{-4}}{\Msun}.
\cite{sieverdingProduction44TiIrongroup2023} has a 3D neutrino-driven supernova of a \unit{18.88}{\Msun}, $Z_\odot$ that yields \unit{\natlog{1.80}{-5}}{\Msun}.
Finally, \cite{wangInsightsProduction44Ti2024} have 18 3D CCSN simulations between \unit{9{-}60}{\Msun} at solar metallicity that produce \unit{\natlog{(0.013{-}2.124)}{-4}}{\Msun} of \titanium[44].

\section{Results}\label{sec:results}

\subsection{Nuclear production of \titanium[44] \textbf{in} O-C shell mergers}\label{sec:results_ti44}

All mixing scenarios considered in this work share a single, common reaction chain that produces \titanium[44] during the \oxygen[]-\carbon[] shell merger: 
\argon[38]$(\alpha,\gamma)$\calcium[42]$(\nt,\gamma)$\calcium[43]$(\pt,\gamma)$\scandium[44]$(\pt,\nt)$\titanium[44]. 
All mixing conditions also share \titanium[44]$(\gamma,\alpha)$\calcium[40] as the dominant destruction channel at the base of the shell where $T=\unit{\natlog{2.77{-}2.83}{9}}{\Kelvin}$. 
\titanium[44]$(\nt,\gamma)$\titanium[45] is also a destruction channel, but to a lesser extent. 

\begin{figure}[!ht]
\centering
\includegraphics[width=\hsize]{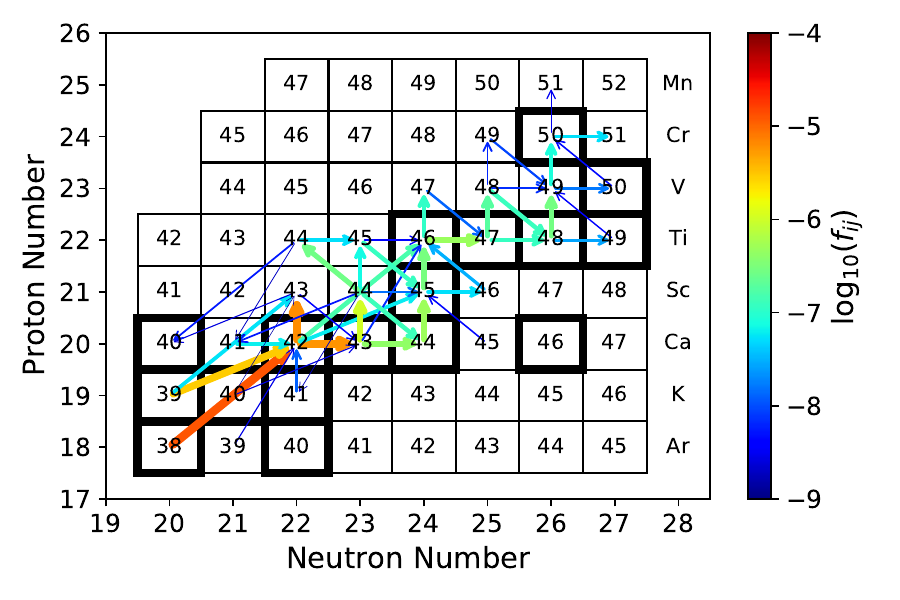}
    \caption{Chart of reactions between isotopes at $m = \unit{1.6}{\Msun}$ $[T = \unit{\natlog{2.494}{9}}{\Kelvin}]$ for the $50\times D_{\mathrm{3D{-}insp.}}$ mixing scenario with an entrainment rate of $\unit{\natlog{4}{-3}}{\Msun \second^{-1}}$ at $t = \unit{110}{\second}$. Both arrow colour and size indicate $\log_{10}(f_{ij})$, the reaction flux as defined in \cite{issaImpact3DMacrophysics2026}, and arrows point in the direction of the reaction.}
\label{fig:reactions}
\end{figure}

The mixing scenarios with a $50\times D_{\mathrm{3D{-}insp.}}$ profile have some additional reactions to this chain.
These scenarios feature significant contributions of \potassium[39]$(\alpha,\pt)$\calcium[42] and \scandium[43]$(\gamma,\pt)$\calcium[42] along with \argon[38]$(\alpha,\gamma)$\calcium[42].
The contribution of \scandium[43] to \calcium[42] may seem counter-intuitive as \Fig{reactions} shows \calcium[42]$(\pt,\gamma)$\scandium[43] as a destruction channel for \calcium[42].
However, it is because the \oxygen[] shell is a convective-reactive enviornment where species can advect on a similar timescale to their reactions.
\Fig{reactions} shows a build-up of \scandium[43] which is mixed deeper in the \oxygen[] and undergoes \scandium[43]$(\gamma,\pt)$\calcium[42] at those hotter temperatures.
\argon[38] in all other scenarios is produced by \silicon[30]$(\alpha,\gamma)$\sulfur[34]$(\alpha,\gamma)$\argon[38], but in the $50\times D_{\mathrm{3D{-}insp.}}$  mixing scenarios it is also produced by \chlorine[35]$(\gamma,\pt)$\sulfur[34]$(\alpha,\gamma)$\argon[38] and \potassium[39]$(\gamma,\pt)$\argon[38].
How \silicon[30] and the light odd-Z isotopes \chlorine[35] and \potassium[39] are produced in the \oxygen[]-\carbon[] shell merger is discussed in \cite{issa3DMacroPhysics2026}.
The importance of the $(\alpha,\gamma)$ channels of \sulfur[34] and \argon[38] for \titanium[44] production are interesting as their $(\pt,\gamma)$ channels may dominate energy production in the \oxygen[]-\carbon[] shell \citep{robertiSPArBurningProton2025}.

\begin{figure}[!ht]
\centering
\includegraphics[width=\hsize]{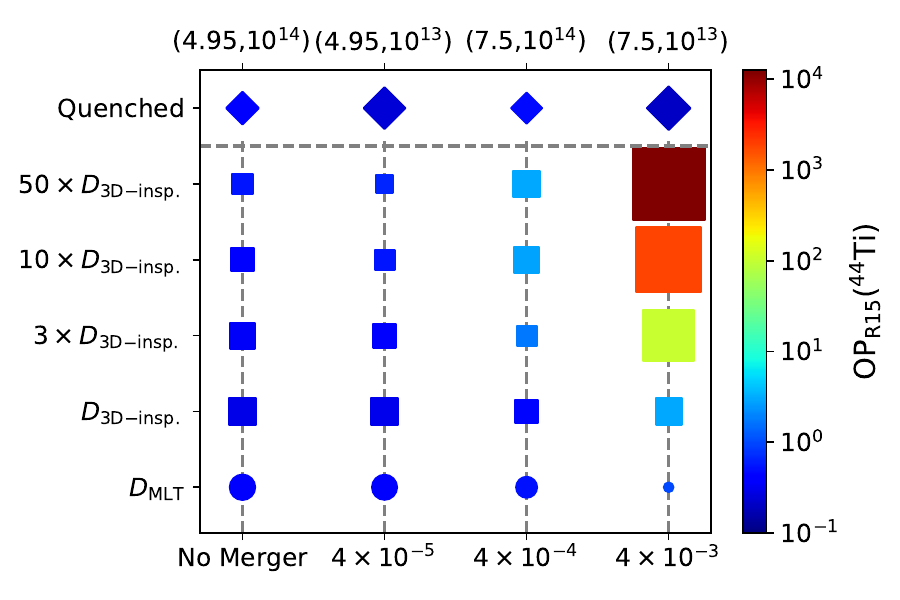}
    \caption{Predicted yields of \titanium[44] compared to the $\unit{15}{\Msun}$ $Z=0.02$ pre-explosive yields ($\OP_{\mathrm{R15}}$). The lower x-axis has the entrainment rates in $\Msun\second^{-1}$ and the upper x-axis has the quenched mixing scenarios summarized as (location of dip in \Mm, maximum extent of the dip in $\cm^{2} \second^{-1}$). Colour indicates magnitude and size indicates distance from $\OP_{\mathrm{R15}}=1$. The explosive production of the $\unit{15}{\Msun}$ $Z=0.02$ model is $\OP_{\mathrm{R15}}=\unit{0.49}{\dex}$.}
\label{fig:OCmerger_Ti44}
\end{figure}

Despite the shared chain of production, \Fig{OCmerger_Ti44} shows \titanium[44] production is dependent on the mixing conditions of the merger to bring sufficient amounts of material to hotter temperatures.
Without entrainment, the \oxygen[] shell does not produce high amounts of \titanium[44].
Entrainment of \carbon[]-shell material supplies a significant source of \calcium[43] to boost production of \scandium[44].
However, the MLT and quenched mixing scenarios show a decrease in \titanium[44] production compared to the pre-explosive model despite high entrainment.
Only models with highest two entrainment rates and a boosted 3D-inspired mixing profile show a significant increase to the production of \titanium[44] compared to the pre-explosive production of the $\unit{15}{\Msun}$, $Z=0.02$ model from~\cite{ritterNuGridStellarData2018}.
In fact, \titanium[44] production is monotonic with mixing speed for the 3D-inspired scenarios.

\begin{figure}[!ht]
\centering
\includegraphics[width=\hsize]{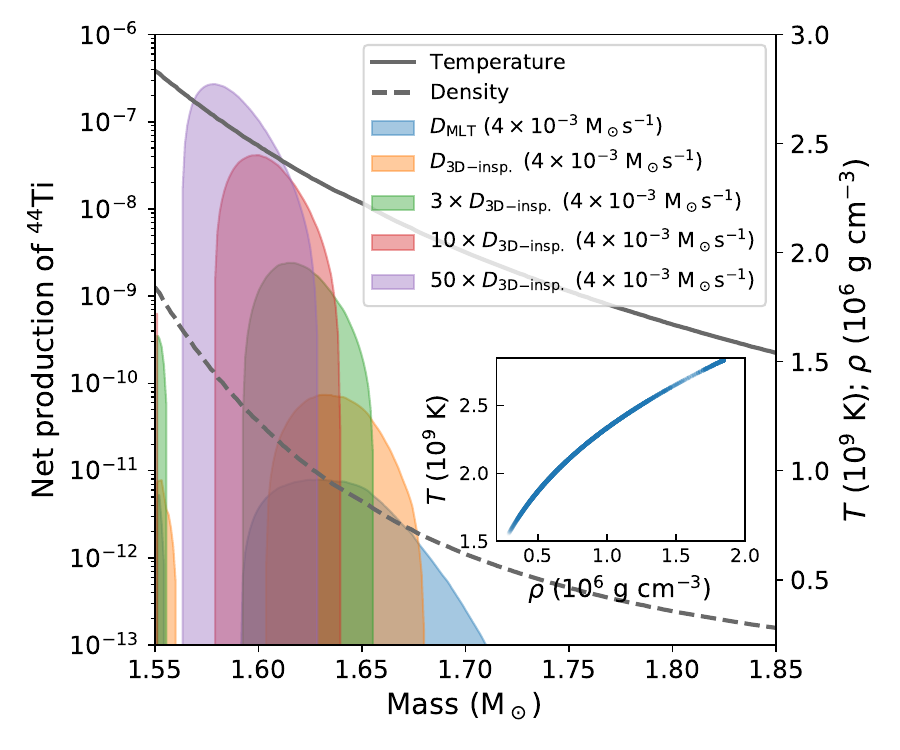}
   \caption{The temperature and density conditions for \titanium[44] production. The coloured shading show the net production for \titanium[44] at each mass coordinate for the $D_\mathrm{MLT}$ and each of the $D_\mathrm{3D{-}insp.}$ mixing profiles with an entrainment rate of $\unit{\natlog{4}{-3}}{\Msun\second^{-1}}$ at $t=\unit{110}{\second}$. The solid and dashed grey lines are the temperature and density. We also include a $\rho$-$T$ inset showing the densities and temperatures where \titanium[44] is produced for all mixing conditions considered in this work.}
\label{fig:rho_t_production}
\end{figure}

Higher mixing speeds are able to mix \calcium[43] and \scandium[44] deeper into the shell more efficiently before producing \titanium[44] that is mixed back up to the cooler top of the \oxygen[] shell.
As mixing speeds increase, the location of peak burning for \calcium[43]$(\pt,\gamma)$\scandium[44] and \scandium[44]$(\pt,\nt)$\titanium[44] move from \unit{1.65}{\Msun} $[T=\unit{\natlog{2.23}{9}}{\Kelvin}]$ in the MLT scenario to \unit{1.59}{\Msun} $[T=\unit{\natlog{2.56}{9}}{\Kelvin}]$ in the $50\times D_{\mathrm{3D{-}insp.}}$ scenario as shown in \Figure{rho_t_production}.
This is a feature of convective-reactive environments because the ratio of mixing and burning timescales changes location as explored in \cite{issaImpact3DMacrophysics2026, issa3DMacroPhysics2026}.

These results demonstrate the importance of both the entrainment rate and mixing speed in for pre-explosive nucleosynthesis of \titanium[44] during an \oxygen[]-\carbon[] shell merger.

\subsection{Mixing and explosive yields of radioactive species}\label{sec:results_mixing_explosive}

\begin{figure*}[]
\centering
\includegraphics[width=\hsize]{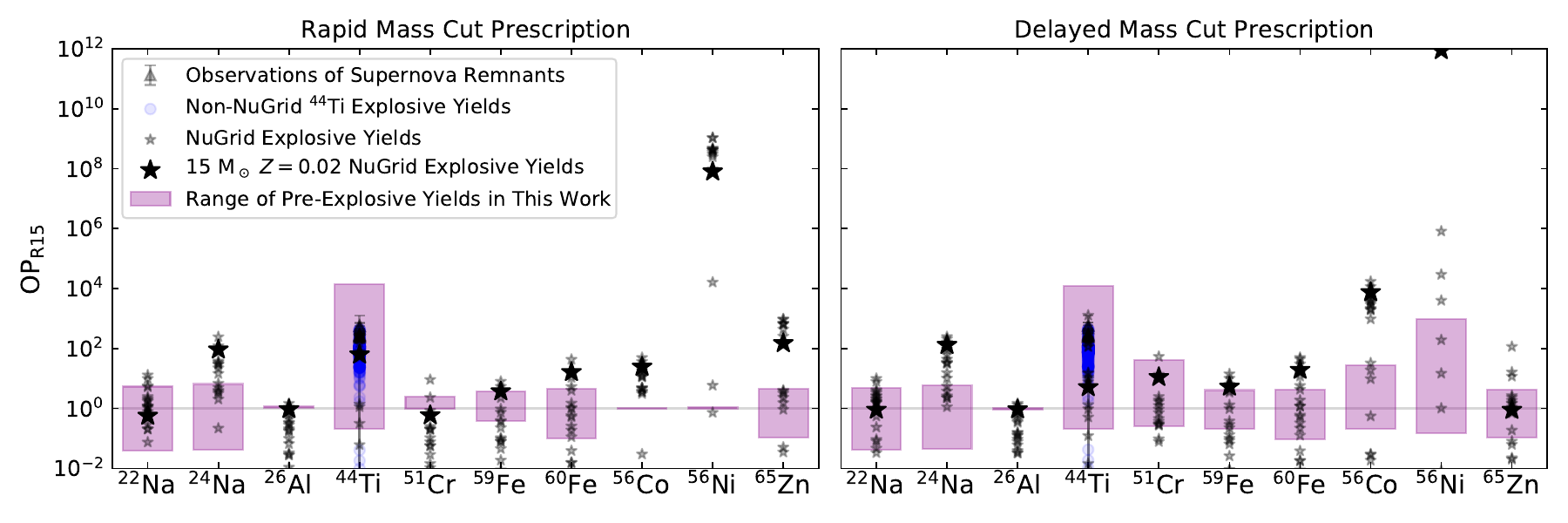}
   \caption{$\OP_\mathrm{R15}$ for the rapid (left) and delayed (right) mass cut prescriptions. We plot (1) inferred yields of \titanium[44] from supernova remnants observations (triangles with error bars, see \Sect{obs_citations} for all citations) (2) explosive \titanium[44] yields of non-NuGrid models (blue circles, see \Sect{model_citations} for all citations) (3) explosive yields of all considered radioactive species from NuGrid models (small black stars) (4) explosive yields of the $\unit{15}{\Msun}$ $Z=0.02$ NuGrid model (large black stars) (5) the maximum and minimum predicted pre-explosive yields for all scenarios (purple bars).}
\label{fig:OCmerger_comp}
\end{figure*}

Mixing impacts our estimated pre-explosive yields for a broader range of radioactive isotopes produced in the \oxygen[] shell beyond \titanium[44].
\Fig{OCmerger_comp} shows the maximum and minimum yields for radioactive isotopes from our pre-explosive mixing scenarios and explosive yields from models and observations for \titanium[44] compared to the pre-explosive yields of the \unit{15}{\Msun} $Z=0.02$ model for both mass cuts.
For this model, the delayed mass cut falls inside of the \oxygen[] shell and the deeper rapid mass cut includes a \silicon[]-burning shell underneath as seen in \Fig{massfrac}.
We apply the mass cut prescription to all NuGrid models and the exploratory yields from this work.
Differences in our pre-explosive mixing yields between the panels of \Fig{OCmerger_comp} demonstrate how the inclusion of the \silicon[]-burning shell modifies the overall contribution from the \oxygen[]-\carbon[] shell merger.

\Fig{OCmerger_comp} show that the mass cut does not play a large role in our yields of \titanium[44].
It also shows that the explosive production of \nickel[56] completely dominates over all mixing scenarios we consider, and hence we do not compare it systematically to models and observations. 
This shows that the \titanium[44] ejecta can have a significant enhancement prior to the explosion without impacting the \nickel[56] yield in 1D models if 3D mixing physics are considered.

\Fig{OCmerger_comp} shows that most radioactive isotopes considered in this work feature a multi-dex spread in their predicted yields, although the range varies with mass cut.
The rapid and delayed mass cuts have average spreads of $\unit{1.54}{\dex}$ and $\unit{2.14}{\dex}$ for our pre-explosive yields respectively (excluding \aluminum[26] which extends to the surface as shown in \Fig{massfrac}).
The rapid mass cut has a noticably decreased mixing impact for \chromium[51], \cobalt[56], and \nickel[56] because these isotopes have a large production in the \silicon[] shell underneath the merger region (\Fig{massfrac}).

Comparing to the explosive yields of the $\unit{15}{\Msun}$, $Z=0.02$ model, these results suggest \sodium[22], \titanium[44], \chromium[51], and \zinc[65] could be dominated by pre-explosive merger nucleosynthesis rather than the explosion.
\iron[59] and \iron[60] could also have significant contributions from the merger.
Among the radioactive isotopes considered, \titanium[44] is the most sensitive to the mixing conditions with a spread of $\unit{4.81}{\dex}$ and $\unit{4.78}{\dex}$ for the rapid and delayed mass cuts.
This, along with \Tab{ti_issa} in \Appendix{appendix_explosive_yields}, clearly shows that mass cut selection does not play a large role in these yields of \titanium[44].

The range of predicted pre-explosive yields in our mixing scenarios for \sodium[22], \titanium[44], and \chromium[51] are comparable or larger than the explosive yields of all massive star models in the NuGrid data set for both mass cuts.
The pre-explosive yields of \iron[59] and \iron[60] are also comparable or larger than many of the explosive yields in the NuGrid data set for both mass cuts.

We have also compared our results to a number of 1D, 2D, and 3D models of massive star yields of \titanium[44] from the literature, some of which include rotation.
\Fig{OCmerger_comp} shows that the range of predicted pre-explosive yields for \titanium[44] can be larger in our 1D models compared to all considered models we compare to.
However, the $\unit{15}{\Msun}$, $Z=0.02$ explosive yields of \nickel[56] are too low compared to 3D simulations of the explosion.
The delayed and rapid mass cuts yield \unit{\natlog{2.55}{-2}}{\Msun} and \unit{\natlog{3.44}{-2}}{\Msun} of \nickel[56], which is $1.5{-}4$ times lower than 3D models of similar mass\citep{wongwathanaratProductionDistribution44Ti2017,sieverdingProduction44TiIrongroup2023,wangInsightsProduction44Ti2024} despite our 3D-inspired models producing similar amounts of \titanium[44].
This makes it clear that even if \titanium[44] produced in a shell merger survives the shock wave, 3D simulations of the explosion are necessary to fully capture nucelosynthesis in supernovae even if the star had a shell merger.

\cite{satoInhomogeneousStellarMixing2025} and \cite{audardChlorinePotassiumEnrichment2025} suggest that Cassiopeia A had an \oxygen[]-\carbon[] shell merger shortly before collapse as an explanation of the measured \magnesium[], \neon[], \silicon[], 
\argon[], \chlorine[] and \potassium[].
Since pre-explosive \titanium[44] production occurs along with these elements in the shell merger and is agnostic to mass cut in our simulations, it points to the ability to use \titanium[44] without a robust \nickel[56] measurement \citep{chieffiSynthesis44Ti56Ni2017}.
Comparing our results to \cite{weinberger44TiEjectaYoung2020}, Cassiopeia A ejected \unit{\natlog{(2.6\pm 0.6)}{-4}}{\Msun} of \titanium[44] which is between our $3\times$ and $10\times D_\mathrm{3D{-}insp.}$ scenarios with the fastest entrainment rate of \unit{\natlog{4}{-3}}{\Msun\second^{-1}}. 
If Cassiopeia A had a shell merger, its \titanium[44] yield could be explained by its \oxygen[] shell having a convective downturn and mixing speeds boosted by $3{-}10$ compared to MLT without much synthesis in the explosion.
While it is unlikely that all of Cassiopeia A's \titanium[44] is pre-explosive, since both \titanium[44] and \nickel[56] are explained by 3D explosions \citep{wongwathanaratProductionDistribution44Ti2017, sieverdingProduction44TiIrongroup2023,wangInsightsProduction44Ti2024}, it is clear that a portion may be pre-explosive if the progenitor had a shell merger.
Finally, Cassiopeia A and SN 1978A exclude our $50\times$ scenario at the fastest entrainment rate since the scenario overproduces \titanium[44].

These results demonstrate the critical importance of understanding the mixing conditions during the \oxygen[]-\carbon[] shell merger for predicting the yields of radioactive isotopes for pre-supernova models, especially \titanium[44].

\section{Conclusions}\label{sec:conclusions}

This work highlights the need to understand the 3D macro physics of mixing in the \oxygen[] shell during a merger.
We have shown how \titanium[44] can be significantly produced by pre-explosive nucleosynthesis during an \oxygen[]-\carbon[] shell merger in massive stars by \scandium[44]$(p,n)$\titanium[44] and its sensitivity to both entrainment and mixing speed.
We have also found that \sodium[22], \chromium[51], \iron[59,60], and \zinc[65] also have relevant pre-explosive nucleosynthesis depending on the mixing conditions for both mass cut prescriptions used.
Further, we have shown for \titanium[44] that the pre-explosive \oxygen[]-\carbon[] shell production can dominate over explosive yields from a wide range of massive star models and are consistent with observations.

These predicted results do not properly account for the pre-collapse evolution of the \oxygen[]-\carbon[] shell merger and how feedback effects may alter the final nucleosynthesis.
However, this work does show the importance of including insights from 3D macro physics into 1D stellar evolution and explosion models to better understand the nucleosynthesis of radioactive isotopes in massive stars.
Another feature missing from this model is a treatment of rotation as \cite{chieffiSynthesis44Ti56Ni2017} found it could boost pre-explosive \titanium[44] production during convective \oxygen[] shell burning.
3D effects of a convective downturn and boosted mixing speeds compared to MLT are general features of \oxygen[] shell convection \citep{jonesIdealizedHydrodynamicSimulations2017}, and our results find a spread even in the absence of a merger (left column of \Fig{OCmerger_Ti44}).
3D explosive effects are also necessary to understand the production of \titanium[44] and \nickel[56] for to match supernova remnants \citep{wongwathanaratProductionDistribution44Ti2017, sieverdingProduction44TiIrongroup2023,wangInsightsProduction44Ti2024}, and should be considered in further works.
The relationship between mergers and the subsequent explosion need further study as the non-radial perturbations from it may also play a significant role in how the subsequent supernova explodes \citep{couchRevivalStalledCorecollapse2013, mullerSupernovaSimulations3D2017, mullerHydrodynamicsCorecollapseSupernovae2020, andrassy3DHydrodynamicSimulations2020,yadavLargescaleMixingViolent2020, fieldsDevelopmentMultidimensionalProgenitor2020, rizzutiShellMergersLate2024a, laplaceItsWrittenMassive2025}.

If the \oxygen[]-\carbon[] merger contributes to the yields of radioactive isotopes of massive stars, then understanding the mixing conditions has a wide range of implications for observations.
Understanding supernova remnants like Cassiopeia A \citep{satoInhomogeneousStellarMixing2025}, pre-solar grain measurements of \calcium[44] from \titanium[44] decays \citep{pignatariPRODUCTIONCARBONRICHPRESOLAR2013,liuPresolarGrainsProbes2024a}, and supernova light curves powered by \sodium[22] or \titanium[44] decays \citep{pignatariProductionRadioactive22Na2025} are all potentially impacted by the mixing discussed in this work.
Upcoming missions such as \textit{COSI} will measure decay lines for \titanium[44] and \iron[60], providing more data to constrain massive star nucleosynthesis with supernova remnants \citep{tomsickComptonSpectrometerImager2019}, although additional measurements such as \nickel[56] will be needed to distinguish progenitor physics. 
We conclude 3D hydrodynamic macro physics are a crucial part of understanding the nucleosynthesis and origin of radioactive isotopes, particularly \titanium[44], during \oxygen[]-\carbon[] mergers.

\begin{acknowledgements}
    We would like to thank Marco Pignatari, Lorenzo Roberti, and Frank Timmes for useful discussions and feedback. FH is supported by a Natural Sciences and Engineering Research Council of Canada (NSERC) Discovery Grant. 
    This research has used the Astrohub online virtual research environment (\url{https://astrohub.uvic.ca}), developed and operated by the Computational Stellar Astrophysics group at the University of Victoria and hosted on the Digital Research Alliance of Canada Arbutus Cloud at the University of Victoria. 
    These simulations were performed on Digital Research Alliance's Niagara supercomputer cluster operated by SciNet at the University of Toronto.
    This work benefited from interactions and workshops co-organized by The Center for Nuclear astrophysics Across Messengers (CeNAM) which is supported by the U.S. Department of Energy, Office of Science, Office of Nuclear Physics, under Award Number DE-SC0023128. 
\end{acknowledgements}

\bibliography{paper}{}
\bibliographystyle{aasjournalv7}

\appendix

\section{\titanium[44] Yields}\label{sec:appendix_explosive_yields}

We provide here tables of our \titanium[44] yields and those of the NuGrid data set for both the delayed and rapid mass cut prescriptions. 
\Tab{ti_ritter} has the pre-explosive and explosive \titanium[44] yields from the massive stars from the NuGrid data set \citep{ritterNuGridStellarData2018}.
\Tab{ti_issa} has the predicted pre-explosive \titanium[44] yields from this work that are calculated according to the method described in \cite{issa3DMacroPhysics2026}.

\begin{table}[h!]
\centering
\caption{Pre-explosive and explosive \titanium[44] yields of the \cite{ritterNuGridStellarData2018} models.}
\begin{tabular}{llcccc c}
\hline
&& \multicolumn{2}{c}{Delayed Yields ($\Msun$)} & \multicolumn{2}{c}{Rapid Yields ($\Msun$)} & \\
\cline{3-4} \cline{5-6}
Mass & Metallicity
& Pre-Explosive & Explosive
& Pre-Explosive & Explosive
& \oxygen[]-\carbon[] Merger \\
\hline
12&0.0001&\natlog{1.09}{-10}&\natlog{1.25}{-4}&\natlog{7.33}{-10}&\natlog{1.48}{-4}&\\
12&0.001&\natlog{1.41}{-11}&\natlog{6.30}{-5}&\natlog{3.23}{-10}&\natlog{1.35}{-4}&Yes\\
12&0.006&\natlog{3.13}{-11}&\natlog{1.09}{-4}&\natlog{5.98}{-10}&\natlog{1.35}{-4}&\\
12&0.01&\natlog{2.05}{-9}&\natlog{1.02}{-4}&\natlog{4.03}{-9}&\natlog{1.35}{-4}&Yes\\
12&0.02&\natlog{2.67}{-11}&\natlog{8.37}{-5}&\natlog{2.99}{-10}&\natlog{1.14}{-4}&\\
\toprule
15&0.0001&\natlog{1.92}{-10}&\natlog{8.39}{-6}&\natlog{8.45}{-10}&\natlog{8.63}{-5}&\\
15&0.001&\natlog{1.74}{-11}&\natlog{4.14}{-6}&\natlog{2.58}{-10}&\natlog{4.54}{-5}&\\
15&0.006&\natlog{2.13}{-10}&\natlog{6.16}{-5}&\natlog{9.92}{-10}&\natlog{1.30}{-4}&\\
15&0.01&\natlog{1.79}{-9}&\natlog{8.03}{-6}&\natlog{1.95}{-9}&\natlog{2.80}{-5}&Yes\\
15&0.02&\natlog{5.40}{-7}&\natlog{2.75}{-6}&\natlog{5.55}{-7}&\natlog{3.50}{-5}&Yes\\
\toprule
20&0.0001&\natlog{3.69}{-12}&\natlog{5.19}{-7}&\natlog{1.02}{-10}&\natlog{1.75}{-7}&\\
20&0.001&\natlog{84}{-11}&\natlog{5.99}{-6}&\natlog{4.28}{-10}&\natlog{1.52}{-4}&\\
20&0.006&\natlog{1.65}{-11}&\natlog{6.92}{-4}&\natlog{5.20}{-10}&\natlog{1.63}{-4}&\\
20&0.01&\natlog{2.77}{-7}&\natlog{2.77}{-7}&\natlog{5.68}{-7}&\natlog{6.30}{-7}&Yes\\
20&0.02&\natlog{2.67}{-11}&\natlog{7.13}{-7}&\natlog{2.72}{-11}&\natlog{7.68}{-7}&\\
\toprule
25&0.0001&\natlog{2.41}{-14}&\natlog{4.64}{-11}&\natlog{6.29}{-14}&\natlog{4.23}{-11}&\\
25&0.001&\natlog{2.16}{-13}&\natlog{6.64}{-8}&\natlog{3.83}{-13}&\natlog{3.40}{-8}&\\
25&0.006&\natlog{4.29}{-13}&\natlog{4.44}{-9}&\natlog{1.00}{-25}&\natlog{4.12}{-18}&\\
25&0.01&\natlog{8.24}{-13}&\natlog{5.21}{-6}&\natlog{1.51}{-98}&\natlog{1.51}{-98}&\\
25&0.02&\natlog{6.59}{-11}&\natlog{7.77}{-9}&$-$&$-$&\\
\toprule
\end{tabular}
\label{tab:ti_ritter}
\end{table}

\begin{table}[h!]
\centering    
\caption{Predicted pre-explosive \titanium[44] yields from the~\cite{issaImpact3DMacrophysics2026,issa3DMacroPhysics2026} models for the delayed and rapid mass cut prescriptions.}
\begin{tabular}{lccc}
\toprule
Mixing Profile&Entrainment Rate ($\Msun~\second^{-1}$)& Delayed Yield ($\Msun$) & Rapid Yield ($\Msun$)\\
\toprule
$D_\mathrm{MLT}$& 0 & \natlog{1.94}{-7} & \natlog{1.96}{-7} \\
$D_\mathrm{MLT}$& \natlog{4}{-5} & \natlog{2.00}{-7} & \natlog{2.02}{-7}\\
$D_\mathrm{MLT}$& \natlog{4}{-4} & \natlog{2.83}{-7} & \natlog{2.93}{-7}\\
$D_\mathrm{MLT}$& \natlog{4}{-3} & \natlog{5.40}{-7} & \natlog{5.73}{-7}\\
\toprule
$D_\mathrm{3D{-}insp.}$& 0 & \natlog{1.61}{-7} & \natlog{1.62}{-7}\\
$D_\mathrm{3D{-}insp.}$& \natlog{4}{-5} & \natlog{1.66}{-7} & \natlog{1.68}{-7}\\
$D_\mathrm{3D{-}insp.}$& \natlog{4}{-4} & \natlog{2.46}{-7} & \natlog{2.54}{-7}\\
$D_\mathrm{3D{-}insp.}$& \natlog{4}{-3} & \natlog{1.60}{-6} & \natlog{1.78}{-6}\\
\toprule
$3\times D_\mathrm{3D{-}insp.}$& 0 & \natlog{1.85}{-7} & \natlog{1.87}{-7}\\
$3\times D_\mathrm{3D{-}insp.}$& \natlog{4}{-5} & \natlog{2.22}{-7} & \natlog{2.25}{-7}\\
$3\times D_\mathrm{3D{-}insp.}$& \natlog{4}{-4} & \natlog{9.54}{-7} & \natlog{9.84}{-7}\\
$3\times D_\mathrm{3D{-}insp.}$& \natlog{4}{-3} & \natlog{5.65}{-5} & \natlog{6.28}{-5}\\
\toprule
$10\times D_\mathrm{3D{-}insp.}$& 0 & \natlog{2.40}{-7} & \natlog{2.42}{-7}\\
$10\times D_\mathrm{3D{-}insp.}$& \natlog{4}{-5} & \natlog{3.03}{-7} & \natlog{3.06}{-7}\\
$10\times D_\mathrm{3D{-}insp.}$& \natlog{4}{-4} & \natlog{1.53}{-6} & \natlog{1.57}{-6}\\
$10\times D_\mathrm{3D{-}insp.}$& \natlog{4}{-3} & \natlog{9.83}{-4} & \natlog{1.10}{-3}\\
\toprule
$50\times D_\mathrm{3D{-}insp.}$& 0 & \natlog{2.96}{-7} & \natlog{2.99}{-7}\\
$50\times D_\mathrm{3D{-}insp.}$& \natlog{4}{-5} & \natlog{3.66}{-7} & \natlog{3.70}{-7}\\
$50\times D_\mathrm{3D{-}insp.}$& \natlog{4}{-4} & \natlog{1.67}{-6} & \natlog{1.71}{-6}\\
$50\times D_\mathrm{3D{-}insp.}$& \natlog{4}{-3} & \natlog{6.69}{-3} & \natlog{7.61}{-3}\\
\toprule
Quenched at $\unit{4.95}{\Mm}$ to $\unit{10^{14}}{\cm^2\second^{-1}}$& \natlog{4}{-3} & \natlog{2.38}{-7} & \natlog{2.53}{-7}\\
Quenched at $\unit{4.95}{\Mm}$ to $\unit{10^{13}}{\cm^2\second^{-1}}$& \natlog{4}{-3} & \natlog{1.29}{-7} & \natlog{1.37}{-7}\\
Quenched at $\unit{7.5}{\Mm}$ to $\unit{10^{14}}{\cm^2\second^{-1}}$& \natlog{4}{-3} & \natlog{2.61}{-7} & \natlog{2.79}{-7}\\
Quenched at $\unit{7.5}{\Mm}$ to $\unit{10^{13}}{\cm^2\second^{-1}}$& \natlog{4}{-3} & \natlog{1.10}{-7} & \natlog{1.19}{-7}\\
\toprule
\end{tabular}
\label{tab:ti_issa}
\end{table}

\end{document}

%% file: abbrev.tex
\newcommand{\n}{\ensuremath{\mem{n}}}

\newcommand{\p}{\ensuremath{\mem{p}}}
\newcommand{\e}{\ensuremath{\mem{e}}}

\newcommand{\h}{\ensuremath{\mem{H}}}

\newcommand{\nine}{\ensuremath{^{59}\mem{Ni}}}

\newcommand{\dex}{\ensuremath{\, \mathrm{dex}}} 
\newcommand{\beq}{\begin{equation}}
\newcommand{\beqa}{\begin{eqnarray}}
\newcommand{\eeq}{\end{equation}}
\newcommand{\eeqa}{\end{eqnarray}}
\newcommand{\bedis}{\begin{displaymath}}
\newcommand{\edis}{\end{displaymath}}

%% file: code.tex
\newcommand{\code}[1]{\texttt{#1}}

\newcommand{\mesa}{\code{MESA}}
\newcommand{\MESA}{\mesa}

\newcommand{\mppnp}{\code{mppnp}} % multi-zone post-processing network parallel

%% file: concepts.tex
\newcommand{\pnucn}{\mbox{$p$ nuclei}}

%% file: derivatives.tex
\newcommand{\D}{{\mathrm d}}

%% file: formatting.tex
\newcommand{\Tabff}[1]{{\ref{tab:#1}}}
\newcommand{\Tab}[1]{{Table~\Tabff{#1}}}

\newcommand{\pan}[1]{{\textit{#1}}}

\newcommand{\FIGFF}[2]{{\ref{fig:#2}\pan{#1}}}

\newcommand{\FIG}[2]{{Fig.~\FIGFF{#1}{#2}}}
\newcommand{\Fig}[1]{{\FIG{}{#1}}}

\newcommand{\Figure}[1]{{Figure~\FIGFF{}{#1}}}

\newcommand{\Sectff}[1]{{\ref{sec:#1}}}
\newcommand{\Sect}[1]{{\S\Sectff{#1}}}
\newcommand{\Appendix}[1]{{Appendix~\Sectff{#1}}}

\newcommand{\isofont}[1]{{\mathrm{#1}}}
\newcommand{\isomass}[1]{{\ensuremath{\isofont{^{#1}}}}}
\newcommand{\isocharge}[1]{{\ensuremath{\isofont{_{#1}}}}}
\newcommand{\isotope}[3]{{\ensuremath{\isocharge{#1}\isomass{#2}\isofont{#3}}}}

\newcommand{\I}[2]{{\isotope{}{#1}{#2}}}

\newcommand{\Ep}[1]{{\ensuremath{10^{#1}}}}
\newcommand{\E}[1]{{\ensuremath{\powersep\Ep{#1}}}}

\newcommand{\powersep}{\times}

\newcommand{\mem}[1]{\ensuremath{\mathrm{ #1}}}

%% file: nuclides.tex
\newcommand{\nuclei}[2]{\ensuremath{\mathrm{^{#1}#2}}}

\newcommand{\neutron}{\ensuremath{n}}
\newcommand{\nt}{\neutron}
\newcommand{\proton}{\ensuremath{p}}
\newcommand{\pt}{\proton}
\newcommand{\hydrogen}[1][1]{\nuclei{#1}{H}}

\newcommand{\carbon}[1][12]{\nuclei{#1}{C}}

\newcommand{\oxygen}[1][16]{\nuclei{#1}{O}}

\newcommand{\neon}[1][20]{\nuclei{#1}{Ne}}
\newcommand{\sodium}[1][23]{\nuclei{#1}{Na}}
\newcommand{\magnesium}[1][24]{\nuclei{#1}{Mg}}
\newcommand{\aluminum}[1][27]{\nuclei{#1}{Al}}
\newcommand{\silicon}[1][28]{\nuclei{#1}{Si}}
\newcommand{\phosphorus}[1][31]{\nuclei{#1}{P}}
\newcommand{\sulfur}[1][32]{\nuclei{#1}{S}}
\newcommand{\chlorine}[1][35]{\nuclei{#1}{Cl}}
\newcommand{\argon}[1][36]{\nuclei{#1}{Ar}}
\newcommand{\potassium}[1][39]{\nuclei{#1}{K}}
\newcommand{\calcium}[1][40]{\nuclei{#1}{Ca}}
\newcommand{\scandium}[1][45]{\nuclei{#1}{Sc}}
\newcommand{\titanium}[1][48]{\nuclei{#1}{Ti}}

\newcommand{\chromium}[1][52]{\nuclei{#1}{Cr}}

\newcommand{\iron}[1][56]{\nuclei{#1}{Fe}}
\newcommand{\cobalt}[1][59]{\nuclei{#1}{Co}}
\newcommand{\nickel}[1][58]{\nuclei{#1}{Ni}}

\newcommand{\zinc}[1][64]{\nuclei{#1}{Zn}}

%% file: units.tex
\newcommand{\numberspace}{\ensuremath{\;}}

\newcommand{\unitstyle}[1]{\ensuremath{\mathrm{#1}}}

\newcommand{\natlog}[2]{\ensuremath{#1\times 10^{#2}}} % a*10^b   

\newcommand{\centi}{\unitstyle{c}}
\newcommand{\kilo}{\unitstyle{k}}
\newcommand{\Mega}{\unitstyle{M}}

\newcommand{\meter}{\unitstyle{m}}

\newcommand{\second}{\unitstyle{s}}
\newcommand{\Kelvin}{\unitstyle{K}}
\newcommand{\K}{\Kelvin}  %degrees Kelvin

\newcommand{\cm}{\centi\meter}

\newcommand{\eV}{\unitstyle{eV}}        %eV
\newcommand{\Msun}{\ensuremath{\unitstyle{M}_\odot}}

\newcommand{\yr}{\unitstyle{yr}}        %year
\newcommand{\Mm}{\Mega\meter}   %megameters
\newcommand{\OP}{\ensuremath{\unitstyle{OP}}} % overproduction

\newcommand{\unit}[2]{\ensuremath{#1\numberspace\mathrm{#2}}}